\documentclass[aps,prl,twocolumn,showpacs,superscriptaddress,groupedaddress]{revtex4}  
\usepackage{graphicx,epsfig}  
\usepackage{dcolumn}   
\usepackage{bm}        
\usepackage{amssymb}   
\usepackage{color}
\usepackage[caption = false]{subfig}
\begin{document}

\title{Experimental observation of a first order phase transition in a complex plasma mono-layer crystal}

\author{M.G. Hariprasad}
\email{hari.prasad@ipr.res.in}

\author{P. Bandyopadhyay}

\author{Garima Arora}

\author{A. Sen}

\affiliation{Institute For Plasma Research, HBNI, Bhat, Gandhinagar, Gujarat, India, 382428}

\begin{abstract}
The formation and melting of a mono-layered charged dust particle crystal in a DC glow discharge Argon plasma is studied. The nature of the melting/formation process is established as a first order phase transition from the nature of the variations in the Coulomb coupling parameter, the dust temperature, the structural order parameter and from the existence of  a hysteresis behavior. Our experimental results are distinctly different from existing theoretical predictions for 2D crystals based on the KTHNY mechanism or the Grain boundary induced melting and indicate a novel mechanism that is akin to a fluctuation induced first order phase transition that has not been observed before in complex plasmas.

\end{abstract}

\maketitle

{\it Introduction.-}
  The phase behaviour of two dimensional structures, particularly the nature of their melting transition, has long been a subject of theoretical and experimental interest and also the source of some controversy. Some commonly studied two dimensional structures are molecular monolayers formed by surfactants spread on a water layer \cite{surfactants}, electrons on the surface of liquid Helium \cite{electron2d}, colloidal suspensions of charged sub-micron spheres \cite{colloid2d} and more recently single layer crystalline structures of charged micro-particles (dust) suspended in the electric sheath of a plasma \cite{melzer,dccrystal,meltingtwod2,plasmacrystal}. Two dimensional structures have also been the subject of many computer simulations based on simple theoretical model systems \cite{simulation1,simulation2,simulation3,vaulina}. The most well known theory for two dimensional melting is the one proposed by KTHNY \cite{kthny} which describes the melting as a two stage transition process with an intermediate hexatic phase. The melting begins by forming dislocations and disclinations in the 2D crystalline structure and as a result the long range translational order breaks down and leads to a hexatic phase. The transitions from the solid to the hexatic and then on to the liquid phase are both continuous in nature constituting second order transitions. Other competing mechanisms of melting consist of the Grain-boundary induced (GBI) melting theory \cite{gbi,gbit}, the density wave theory \cite{densitywave} and the instability triggered theory \cite{instability}. The GBI  melting \cite{gbi,gbit} proceeds through long arrays of dislocations at the lattice boundaries that drive a first order phase transition from the crystalline phase directly to the liquid phase. Density wave theory \cite{densitywave} deals with the change of entropy and structure factor variations, but it fails to establish the order of the phase transition. Likewise simulation studies to date do not provide a definitive picture of the nature of the 2D melting phase transition and the question still remains open.\par
The advent of dusty plasma crystals in recent times has provided a strong impetus to the experimental study of 2D melting since the process can be well diagnosed using non-perturbative techniques and a number of such studies have addressed this problem \cite{melzer,gbinosenko,modeinstability,dustacousticinstability,meltingtwod2,plasmacrystal}. In one of the earliest studies, Melzer \textit{et al.} \cite{melzer} reported that their experimental findings on phase transition could best be explained by incorporating features from both the KTHNY theory and the GBI melting theory. They observed that the transition was of a continuous second order nature with respect to a change in the neutral gas pressure \cite{melzer}. GBI melting was reported experimentally by Nosenko {\it et al.} \cite{gbinosenko} in equilibrium and non-equilibrium regimes by laser heating of the dust crystal and they showed that the crystal defect concentration followed a power-law scaling with the dust temperature. A host of other works have associated the anomalous heating of the dust particles and the concomitant melting of the crystal with the onset of various instabilities in the system \cite{modeinstability,dustacousticinstability}. Couedel {\it et al.} \cite{modeinstability} identified the onset of a mode-coupling instability due to the resonant interaction of dust lattice modes as the cause for the dust heating and melting of the crystal. No mention is made about the nature of the phase transition. In the experiment by Sheridan \cite{dustacousticinstability} a dust acoustic instability was induced through externally driven vertical oscillations of the dust particles. The instability induced heating of the dust crystal created lattice defects leading to a hexatic phase before the onset of melting. The order of the phase transition was not explicitly ascertained but indirectly conjectured to be continuous.  A second order melting of a multi-layer dusty plasma crystal was also observed by Schweigert {\it et al.} \cite{wake} who attributed the heating to an ion streaming induced instability caused by the presence of a second layer of dust particles below the primary crystal layer. None of the above experiments provide a clear evidence of the existence of a first order phase transition for a 2D dusty plasma crystal. Although Nosenko {\it et al.} \cite{gbinosenko} claim that their observed phase transition should be of first order since the melting scenario appears to follow the GBI theory they provide no detailed analysis on the nature of the transition. Analytical studies accompanied by kinetic simulations by Joyce {\it et al.} \cite{instability} predict a first order phase transition for instability (in this case an ion flow induced two stream instability) triggered melting but there is yet no experimental verification of such a conjecture. \par 
In this paper we report our experimental observation on the existence of a first order phase transition in the melting of a two dimensional dusty plasma crystal created in a DC glow discharge plasma. The experimental signatures   supporting this claim are as follows. The Coulomb dust crystal is found to undergo an abrupt transition into the liquid state when the background neutral pressure is  decreased slightly (by less than 0.2 Pa) beyond a critical pressure of $P_c =6.9 Pa $. At this transition point the dust temperature shows a steep rise of nearly 25 times the equilibrium temperature and the Coulomb coupling parameter also displays a sharp change. Similar behaviour is also observed in the variation of the pair correlation function and the structural order parameter. An additional experimental signature that lends further support to a first order phase transition is the existence of hysteresis when the pressure change is reversed to refreeze the crystal. Furthermore, the excitation of an unstable mode is identified in the liquid phase after the phase transition.  The dynamical origin of the oscillation is not definitively known but is most likely due to the non-reciprocal wake forces from a few particles that get vertically displaced from the crystal during an early compressive phase. Hence the observed melting phenomenon closely resembles a fluctuation induced first order phase transition as has been conjectured in some past studies in other media \cite{fluctuation1,fluctuation2}.\par

\begin{figure}
\includegraphics[scale=0.4]{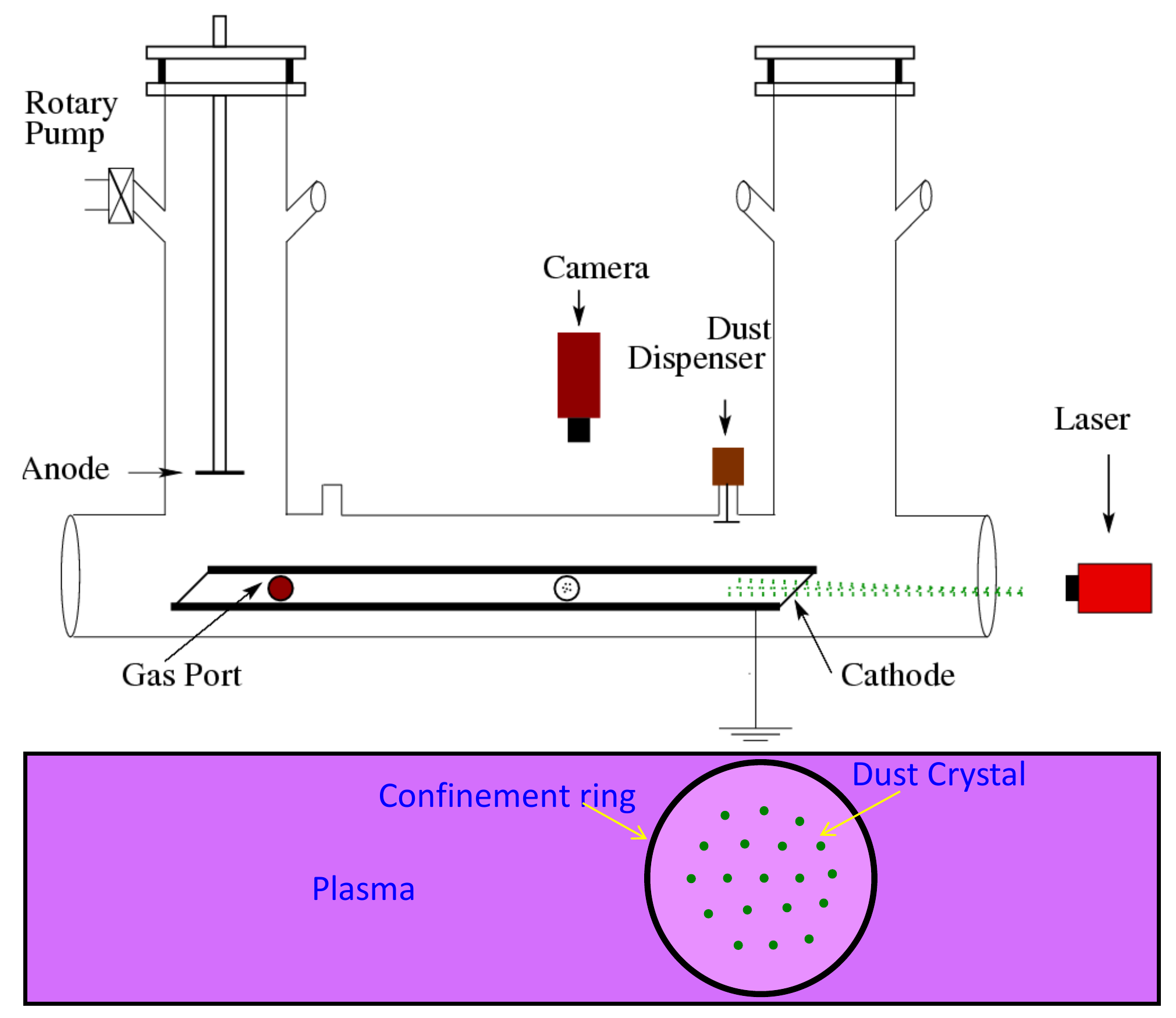}
\caption{\label{fig:fig0} Experimental Setup}
\end{figure}

{\it Experiment.-}
An inverted $\Pi$-shaped Dusty Plasma Experimental  (DPEx) device, as shown in Fig.~\ref{fig:fig0}, is used to carry out the present set of experiments \cite{surabhi,dccrystal}. A rotary pump is connected to one of the arms of the device to evacuate the chamber to attain a base pressure of $p=0.1$ ~ Pa. A long tray shaped grounded cathode and a disc shaped circular anode are used to produce the plasma.  A dust dispenser is mounted on the connecting tube to dispense melamine formaldehyde (MF) mono-dispersive spherical dust particles into the chamber. Just under the dispenser, a metal circular ring is placed on the cathode to confine the dust particles in the plasma environment. A green laser and a CCD camera are used to capture the dynamics of the dust particles.\par
To begin with, the experimental vessel is pumped down to the base pressure and Argon gas is then flushed several times using a mass flow controller. The chamber is pumped down to the base pressure every time for the purpose of removing the impurities from the chamber. Finally, the filling pressure is set in the range of 3-10 Pa by adjusting the pumping speed and the gas flow rate. The glow discharge Argon plasma is then initiated by applying a DC voltage in the range 300-600 V between the anode and the cathode at which the plasma current varies from 5-15 mA. Mono-dispersive spherical dust particles of melamine formaldehyde (MF) with a diameter 10.66 $\mu$m are then introduced into the plasma by shaking the dust dispenser. These particles get negatively charged in the plasma environment, and levitate in the cathode sheath region where the electrostatic force acting on the particles balances the gravitational force. The cathode sheath is basically a non-neutral region around the cathode in which there always exists a strong electric field due to the presence of excess amount of ions. This space charge region is the interface region between the wall and the bulk plasma region. In the sheath the charge neutrality condition does not hold. The electrons having a velocity more than $\sqrt{2e\phi(x)/m_e}$ ($\phi(x)$ is being the sheath potential at a particular location $x$) can enter into the sheath region whereas the ions penetrate into the sheath with a velocity higher than the Bohm velocity ($V_B=\sqrt{K_BT_e/m_i}$). The thermal ions in the bulk plasma gain the Bohm velocity while traveling through the pre-sheath region and as a result these high energetic ions heat up the dust particles through Coulomb collisions and ion wake induced instabilities. Later, the dust particles form a circular mono-layer that is maintained by a balance between the  downward gravitational force and the upward force of the cathode sheath electric field.  The circular shape is induced by a metal ring placed on the cathode just below the dispenser \cite{dccrystal}.  A green laser is used to illuminate the particles either in the horizontal or vertical plane and the  Mie-scattered light from the dust particles is captured by a CCD camera with a frame rate of 70 frames/sec, and the data is subsequently analyzed by a particle tracking code to obtain the particle dynamics.\par
\begin{figure}[!hb]
\includegraphics[scale=0.36]{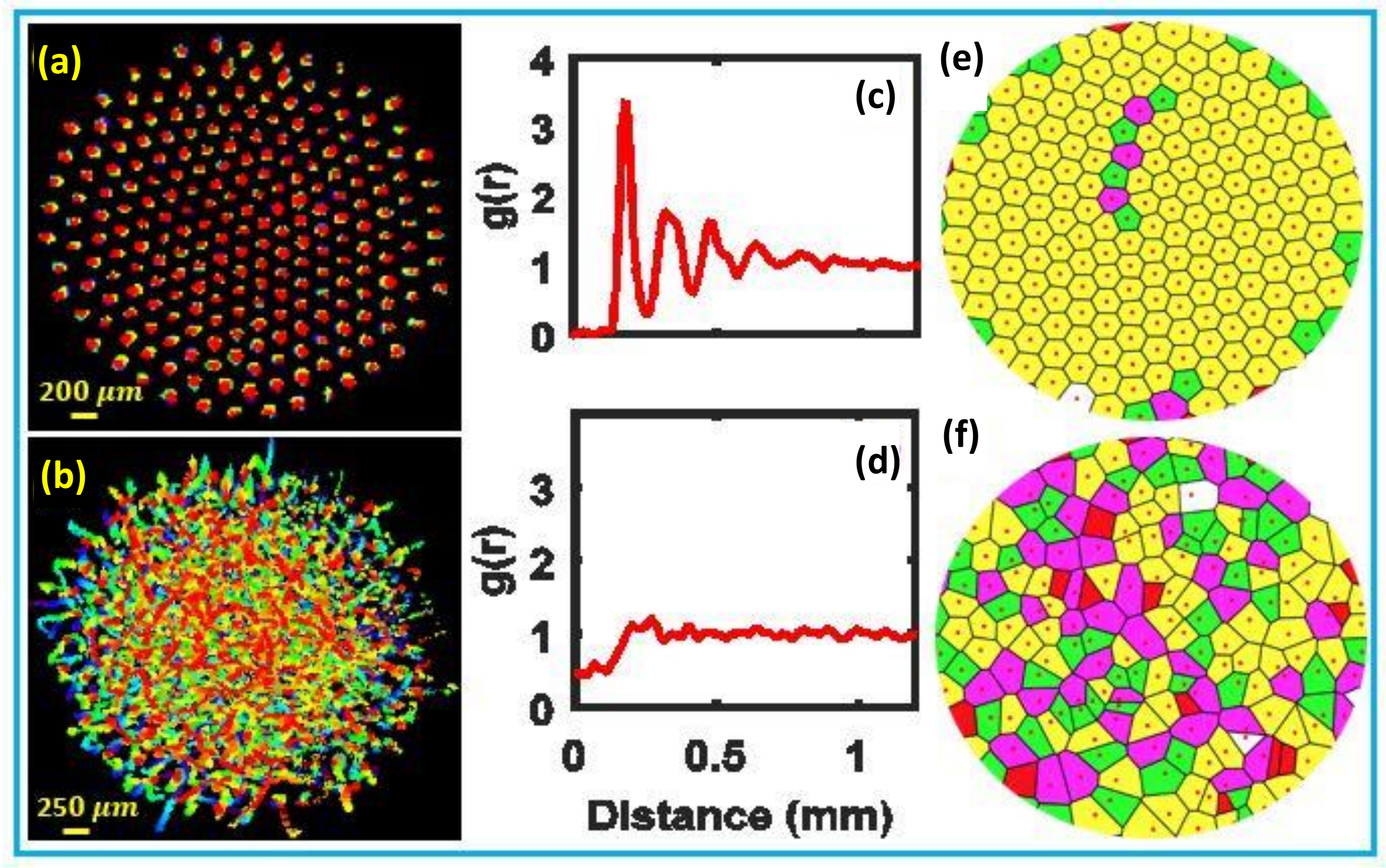}
\caption{\label{fig:fig2} (a), (b) Overlapped position coordinates of dust particles for consecutive 50 frames. (c), (d) Correlation functions and  (e), (f)  Voronoi diagrams corresponds to (a) and (b), respectively. Fig.~\ref{fig:fig2}(a), (c) and (e) correspond to P=6.9 Pa whereas Fig.~\ref{fig:fig2}(b), (d) and (f) are correspond to P=6.7 Pa} 
\end{figure}
{\it Results.-}
The experiments were carried out at a discharge voltage of 480 V with the neutral gas pressure varying from $p=7.5$~Pa to $p=5.5$~Pa at intervals of $0.1 - 0.2$ ~Pa. For $p=7.5-6.7$~Pa, the 2D dust cloud forms an ordered state with hexagonal symmetry as depicted in Fig.~\ref{fig:fig2}(a) which displays the overlapped position coordinates of dust particles for consecutive 50 frames at a neutral pressure of 6.9 Pa . When the neutral pressure is decreased below 6.9~Pa, the crystal disintegrates into a disordered structure of a  highly random nature as shown in Fig.~\ref{fig:fig2}(b). Interestingly, this sudden change happens with a very negligible variation in the neutral pressure from 6.9 Pa to 6.7 Pa.
\begin{figure*}
\includegraphics[scale=1.0]{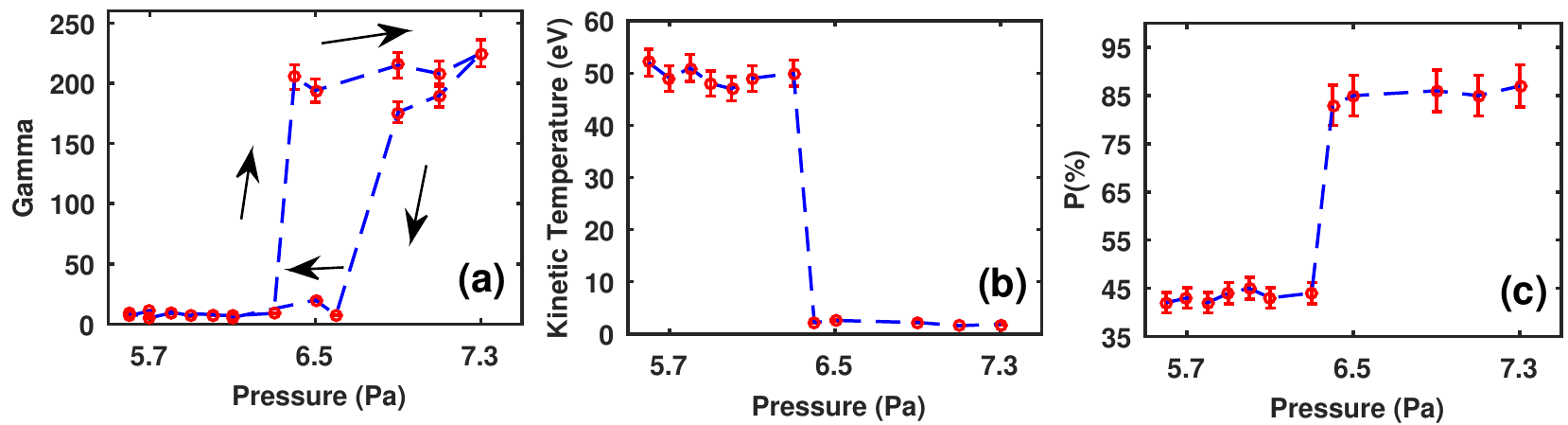}
\caption{\label{fig:fig3}Variations of (a) Coupling parameter   (b) Dust kinetic temperature and (c) Structural Order parameter with neutral gas pressure}
\end{figure*}
The nature of this phase transition was further examined by analyzing the behavior of a number of system and statistical parameters and by employing configuration tools like the Voronoi diagram.\par

A comparison of the nature of the pair correlation function, $g(r)$ \cite{paircorrelation,dccrystal}, for the two states can be seen from Figs~ \ref{fig:fig2}(c) and \ref{fig:fig2}(d).The radial pair correlation function (RPDF) \cite{paircorrelation}, $g(r)$, is used to find the configurational order of a system. This is a measure of the probability of finding particles in the vicinity of a reference particle in the system and provides useful information regarding the structural and thermodynamical properties of a system which is useful to identify the phase state of that system. In Fig.~\ref{fig:fig2}(c), which represents the ordered state of Fig.~\ref{fig:fig2}(a), the pair correlation function is seen to exhibit a number of periodic peaks, which is essentially a signature of a crystalline state. When the pressure is slightly reduced to 6.7 Pa the periodic peaks in the pair correlation function disappear suddenly indicating a sharp phase transition to a disordered fluid state (Fig.~\ref{fig:fig2}(d)). Furthermore the reduction in $g(r)_{max}$ ( where $g(r)_{max}$ is an indirect measure of the Coulomb coupling parameter), also indicates the onset of a phase transition \cite{gamma} from the crystalline state to a liquid state when the pressure is reduced from 6.9~Pa to 6.7~Pa. Fig.~\ref{fig:fig2}(e) and (f) show the Voronoi diagrams for the pressure values of 6.9~Pa and 6.7~Pa, respectively. The hexagonal structures are marked in yellow and the red, pink, green, \textit{etc.} structures correspond to other polygons. At 6.9~Pa (see Fig.~\ref{fig:fig2}(e)), hexagonal structures are found throughout the Voronoi diagram which confirms the findings of the pair correlation analysis. At 6.7 Pa (after the sudden structural transition), the Voronoi diagram shows a complete breakdown of hexagonal symmetry in the system and a number of five-fold and seven-fold cells are formed throughout the dust structure. It clearly signifies the disorder in the dusty plasma system by indicating the absence of both long range and short range orders in the system. Hence, the outcomes from the structural analyses using pair correlation function and the Voronoi diagram essentially indicate that the hexagonal ordering of the dusty plasma structure is spontaneously destroyed indicating a sharp crystal to liquid phase transition. \par

We have next used Langevin's dynamics \cite{coupling,dccrystal} to estimate the screened Coulomb coupling parameter ($\Gamma $) from the displacement distribution functions \cite{coupling,dccrystal}.  Coulomb coupling parameter ($\Gamma $) is defined as the ratio of potential energy between the dust particles to the thermal energy. In a dusty plasma, it can be expressed as $\Gamma=\frac{1}{4\pi\epsilon_0}\frac{Q_d^2f(\kappa)}{K_BT_da} $, where $f(\kappa)$ is $3exp(-\kappa)(1+\kappa+\kappa^2)$ for a 2D Yukawa system, where $Q_d,  K_B, T_d, a $ are the dust charge, Boltzmann constant, dust kinetic temperature, inter particle distance, respectively whereas $\kappa$ is the ratio of inter particle distance to the Debye length. Such an analysis also provides information on the kinetic temperature ($T_d$) of the dust particles that can be extracted from the velocity distribution function. Fig.~\ref{fig:fig3}(a) shows the variation of $\Gamma$ with the neutral gas pressure. At high neutral pressures (beyond $p=6.9$~Pa), the coupling parameter is estimated around $\sim$ 200 and hence the dusty plasma system remains at crystalline state \cite{ikezi1986} over this range of neutral gas pressure. An abrupt decrease in $\Gamma$ from 200 to 8 is found when the gas pressure is reduced from 6.9~Pa to 6.7~Pa, which essentially demonstrates the phase transition of a dusty plasma crystal to a dusty plasma liquid \cite{ikezi1986,vaulina}. It is worth mentioning that the coupling parameter becomes almost twenty-five fold with a pressure variation of only 0.2 ~Pa, which is one of the salient features of a first order phase transition. With further decrease of neutral gas pressure (from 6.7 Pa to 5.5 Pa), the coupling parameter remains almost constant as depicted in Fig.~\ref{fig:fig3}(a). Vaulina \textit{et al.} determined the phase state (whether liquid, hexatic or crystalline) of a Yukawa system by estimating the coupling parameter in their simulation study \cite{vaulina}. According to their findings, the dust system remains in liquid state when $\Gamma $ is less than 98, the hexatic phase is between 98 and 154 and finally the crystalline state happens when $\Gamma $ is greater than 154. In our experiments, the system directly undergoes a transition from the crystal to a liquid state without an intermediate hexatic phase in contrast to the theoretical predictions of the KTHNY theory \cite{kthny}. The order of a phase transition is usually determined from the nature of the change of various thermodynamic parameters. Our use of the Coulomb coupling parameter for such a purpose is justified as it intrinsically includes the variation in the specific heat capacity, the entropy and the internal energy as demonstrated in the simulation studies of Vaulina \textit{et al.} \cite{vaulina}.  Another signature of the sharp first order transition is seen in the variation of the dust temperature with pressure as shown in Fig.~\ref{fig:fig3}(b). A sudden rise in dust temperature from 2 eV to 50 eV is observed during the transition from the crystal phase to the liquid phase. The possible underlying mechanism of this instantaneous heating will be discussed shortly. \par
Furthermore, based on the Voronoi diagram analysis \cite{voronoi}, one can define a structural order parameter 
$P = \frac{N_H}{N_T}*100$, where $N_H$ and $N_T$ are the number of hexagonal structures and the total number of polygons in the Voronoi diagram, respectively. So the higher the number of hexagons, the higher will be the order parameter causing the dust cluster to approach an ordered structure. Fig.~\ref{fig:fig3}(c) shows the structure parameter to experience a jump from $P=85\%$ to $P=40\%$ during the phase transition. It shows the instantaneous generation of five-fold, seven-fold symmetries along with different kinds of defects in the structure. Hence, the findings from all of the above analyses, namely, the pair correlation function, the Voronoi diagram, the Coulomb coupling parameter, the dust temperature and structural order parameter strongly indicate the existence of a first order phase transition of the dusty plasma from a crystalline state to a liquid state due to a small change in the background pressure.\par

To further ascertain the first order nature of the phase transition, another set of experiments was carried out in which the pressure parameter was reversed after the transition to the liquid state in order to refreeze to a crystal state. The refreezing was found not to occur at the same equilibrium pressure but to exhibit a hysteresis behavior.  Such a hysteresis behavior of the Coulomb coupling parameter with respect to the neutral gas pressure is plotted in Fig.~\ref{fig:fig3}(a) where it is seen that the dust cluster initially gets frozen (liquid to solid) at $p=6.9$~Pa in the rising phase of the neutral gas pressure but the same dust cluster is melted back at $p=6.6$~Pa when the neutral pressure is reduced to its initial value. These sharp changes from liquid to crystal and crystal to liquid phases transition are observed at $6.7-6.9$ and $6.8-6.6$ Pa, respectively.  Such a hysteresis phenomenon is a typical signature of first order phase transitions and provides further support to our conclusions arrived at from other experimental analyses. 

\begin{figure}[!hb]
\centering
\includegraphics[scale=0.58]{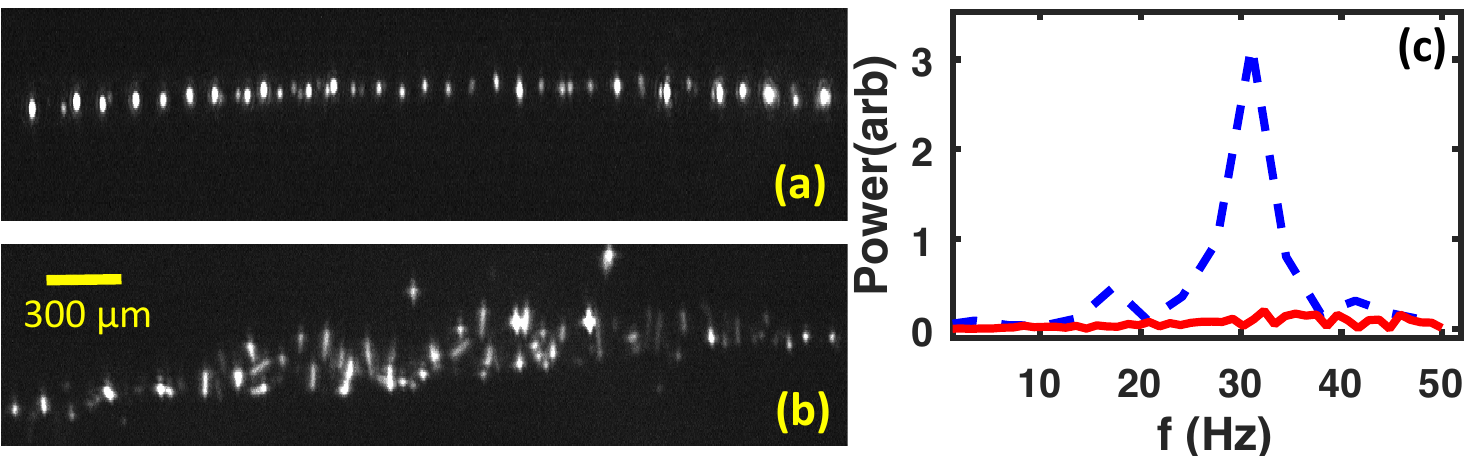}
\caption{\label{fig:fig6} Side images of (a) Crystalline state (b) Liquid state at neutral pressures of 6.9 Pa and 6.7 Pa, respectively. (c) Power spectra of the particle fluctuations of (a) shown as a solid curve and of (b) shown as a dashed curve. }
\end{figure}
\begin{figure}[!hb]
\centering
\includegraphics[scale=0.36]{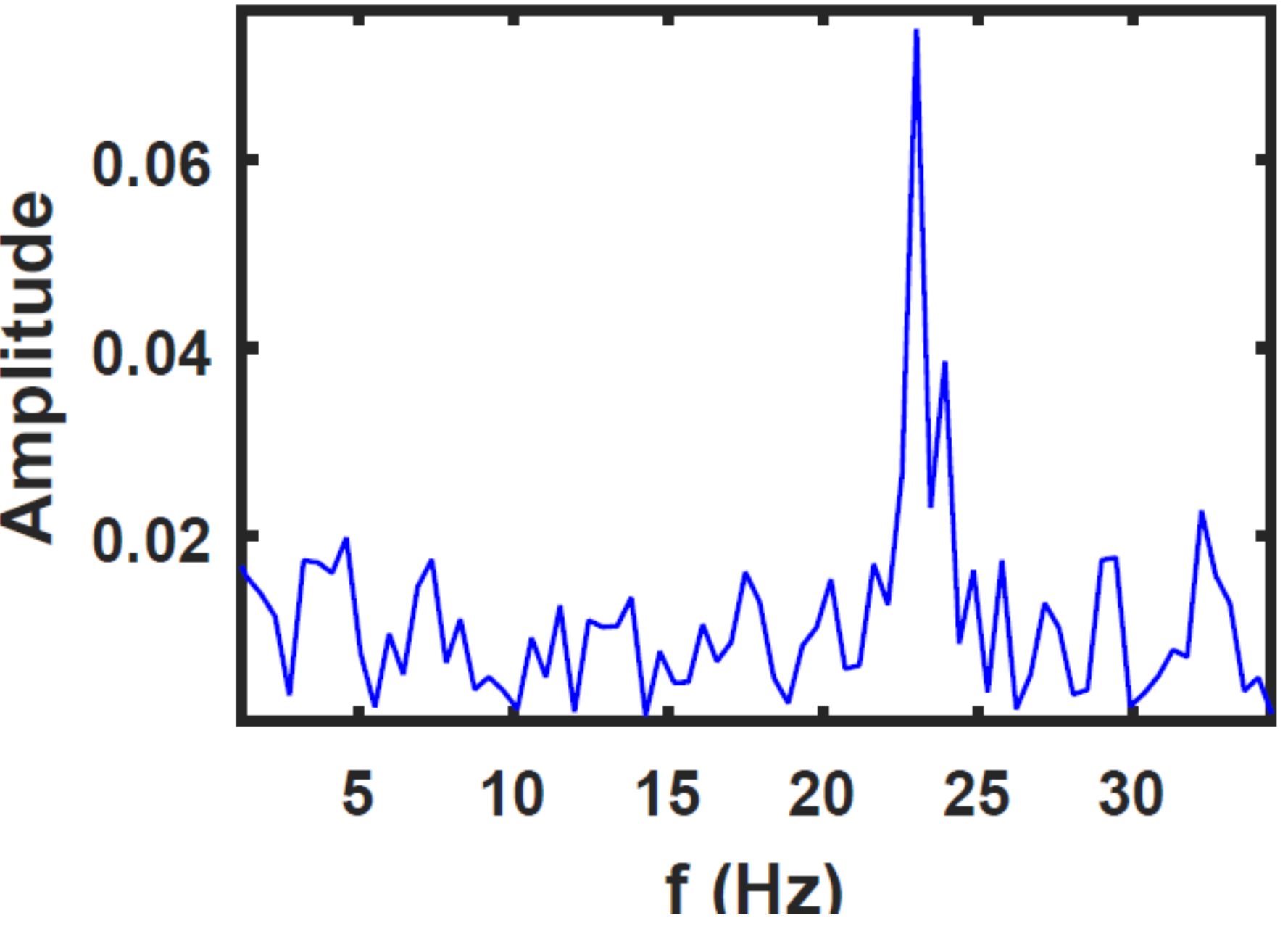}
\caption{\label{fig:fig4} In-plane fluctuation spectra in crystal state just before melting. } 
\end{figure}

To further explore the dynamics of the dust particles during the phase transition, 
we switched the orientation of the laser to illuminate the particles in the plane vertical to the plane of the crystal and the camera was mounted in such a way that it captured only the images in this plane. Some typical images in this configuration for two different cases are presented in Fig.~\ref{fig:fig6}. {In the crystalline state as shown in Fig.~\ref{fig:fig6}(a), the dust particles are aligned in one line and correspond to a mono-layer of a dust crystal whereas at lower pressure large vertical fluctuations set in as is evident from Fig. \ref{fig:fig6}(b). When the neutral gas pressure is reduced to a small value below a threshold pressure $p=6.9$~Pa, the plasma crystal gets compressed (as recently reported by Hariprasad \textit{et al.} \cite{dccrystal}) and as a result some particles get displaced vertically which in turn heats the crystal and the dust temperature rises from 2~ev to 50~ev. This large heating due to the vertically displaced particles can be attributed to the formation of ion wake structures beneath the crystal layer.} The formation of an ion wake resulting in a non-reciprocal force can then trigger an instability in the system as has been predicted in the past \cite{ichiki,wake}. Ichiki \textit{et.al} \cite{ichiki} have demonstrated both experimentally and through simulation studies the formation of a hot crystal and a continuous two-step melting occuring due to the presence of ion wakes. By contrast, in our experiment, the anomalous heating of the crystal and its melting appears to occur in one single step. One significant difference between our experiment and that of Ichiki \textit{et.al} \cite{ichiki} is that in their experiment there always existed a large dust particle beneath the monolayer right from the beginning and hence gave rise to a continuous heating from the wake effects. {In our experiments, a mono-layer of dust particles forms at a higher pressure which then suddenly breaks up due to fluctuations at the lower neutral gas  pressure. The sudden onset of vertical fluctuations seems to be the key reason for the single step heating and first order phase transition observed in our experiments.}\par
In order to verify the presence of an unstable mode we have also investigated the mode spectra of the vertical oscillations in both the crystalline and the liquid states using the side images as shown in Fig.~\ref{fig:fig6}(c). As can be clearly seen the power spectrum for the crystalline state is quite broad and flat (solid curve) indicating the thermal fluctuations of the particles whereas the liquid state shows a large peak corresponding to large amplitude vertical oscillations. The existence of such an oscillation and its experimental coincidence with the crystal melting bears close resemblance to the onset of fluctuation induced first order phase transitions reported in other media like isotropic di-block co-polymers \cite{fluctuation1}, an-isotropic graphite \cite{fluctuation2}, etc. and indicates the availability of a new phase transition channel for two dimensional Coulomb crystals. To further establish the unique nature of this transition channel we have also looked for any evidence of some of the other mechanisms of melting such as the Mode Coupling Instability (MCI) that were invoked in earlier studies \cite{modeinstability1,mciflame,mciivlev,mciMeyer}. One of the distinct features of the MCI is the existence of  mixed polarization wave modes (hybrid modes) arising due to the coupling between the transverse (vertical) modes and the in-plane longitudinal  modes. A physical consequence of such a coupling is that traces (signatures) of the vertical mode can be detected in the in-plane fluctuation spectrum, namely the phonon spectrum. A detailed phonon spectrum of a dusty crystal was obtained by Nunomura {\it et al} \cite{nunomura2002} and displayed in the form of dispersive curves within a Brillouin zone. For our present objective we do not need such detailed dispersive curves; rather a simple and convenient way to identify the longitudinal oscillations is by looking at the frequency power spectrum of the in-plane fluctuations in the particle positions. The signature of an MCI process (hybrid modes) would then show up in that spectrum by the presence of a peak corresponding to the vertical frequency. We have looked for such a signature by constructing the in-plane fluctuation spectrum from our experimental data of the crystalline state just before melting and this is shown in Fig.~\ref{fig:fig4}.  As can be seen that there is no signature of the unstable vertical mode of 30 Hz frequency (see Fig.~\ref{fig:fig6}(c)) in the in-plane spectra. One only sees the presence of an in-plane mode of 24 Hz as shown in Fig.~\ref{fig:fig4}. In addition, the melting is found to be initiated all over the crystal unlike the MCI melting in which the melting initiates from the centre (high density region) and then leads to a flame propagation which melts the full crystal over time \cite{mciflame}. \par
{\it Conclusions.-}
In conclusion, a first order phase transition in a two dimensional dusty plasma crystal has been experimentally observed and validated using a number of diagnostic tests. These include analysis of the nature of changes in the structural order parameter, the Coulomb coupling parameter and the dust temperature. All of these experience sharp changes with a negligible change of neutral gas pressure from a critical pressure value. Additional support is provided by the existence of hysteresis in the transition process. The phase transition is also accompanied by the onset of large vertical oscillations which ostensibly act as the source for the anomalous heating and melting of the crystal. Our results suggest a novel mechanism for a solid to liquid phase transition in a two dimensional dusty plasma crystal that has not been observed before and that has strong similarities with fluctuation induced transitions conjectured in other two dimensional systems. Our findings can be the basis for future experimental,  theoretical modeling and numerical simulation studies of this mechanism that can not only provide rich insights into the behavior of strongly coupled two dimensional plasma systems but also help develop useful linkages with the dynamics of a wider set of two dimensional crystalline systems.\par

{\it Acknowledgement.-} AS is thankful to the Indian National Science Academy
(INSA) for their support under the INSA Senior Scientist Fellowship
scheme.

\bibliography{refer}

\end{document}